\documentclass[12pt, epsfig]{article}
\usepackage[dvips]{color}
\usepackage{epsfig}%
\usepackage[active]{srcltx}%
\usepackage{amsmath,amsfonts,amssymb,amsthm,amstext,amscd,eucal,srcltx}
\usepackage{epsfig,graphicx,bm}
\usepackage{epstopdf, epsf}
\usepackage{dcolumn}
\usepackage{hyperref}
\usepackage{subfigure}
\usepackage{textcomp}

\definecolor{darkgreen}{rgb}{0,0.3,0}
\definecolor{darkblue}{rgb}{0,0,0.3}
\definecolor{darkred}{rgb}{0.7,0,0}

\newcommand{\be}{\begin{equation}}
\newcommand{\bse}{\begin{subequations}}
\newcommand{\ese}{\end{subequations}}
\newcommand{\bea}{\begin{eqnarray}}
\newcommand{\eea}{\end{eqnarray}}
\newcommand{\ba}{\begin{array}}
\newcommand{\ea}{\end{array}}
\newcommand{\ee}{\end{equation}}

\makeatletter \@addtoreset{equation}{section}

\def\dre_g{\delta\rho_g}
\def\dpe_g{\delta P_g}
\def\dqe_g{\delta q_g}
\def\dre{\delta\rho}
\def\dpe{\delta P}
\def\dqe{\delta q}

\def\YM1{\frac{\dot\phi^2}{a^2}}
\def\YM2{\frac{g^2\phi^4}{a^4}}

\begin{document}
\title { Vector inflation by kinetic coupled gravity}

\author{F. Darabi\thanks{Email: f.darabi@azaruniv.edu}, A. Parsiya\thanks{Email: a.parsiya@azaruniv.edu}\\
{\small Department of Physics, Azarbaijan Shahid Madani University, Tabriz 53714-161, Iran.}}

\maketitle

\begin{abstract}
Vector inflation is a newly established model where inflation is driven by non-minimally coupled massive vector fields with a potential term. This model is similar to the model of chaotic inflation with scalar fields, except that for vector fields the isotropy of expansion is achieved either by considering a triplet of orthogonal vector fields or $N$ randomly oriented independent vector fields.
We introduce a new version of vector inflation where the vector field has no potential term but is non-minimally coupled to gravity through the kinetic term. The non-minimal coupling is established by introducing the Einstein tensor besides the metric tensor within the kinetic term of the vector field.
\end{abstract}
\maketitle

\section{Introduction}

Inflationary cosmology is a well known scenario to remedy the problems of
standard cosmology \cite{Inflation-Books, Weinberg}. This scenario can also reproduce, in a very successful way, the current cosmological data through the $\Lambda$CDM model \cite{Bassett-review}. The most important feature
of inflation is the existence of a scalar field which is subjected to the {\it slow-roll} regime, where the kinetic term of the scalar field remains sufficiently small compared to its potential term. The universe undergoes an inflationary stage of expansion in the slow-roll regime, and when the kinetic term becomes comparable to its potential term the inflation is finished. In general, inflationary models benefit of single or multi-scalar field theories. However, in recent years a considerable amount of works have been focused
to model the inflationary scenario for vector fields \cite{vector-inflation}. 
On the other hand, in order to justify the current acceleration of the universe, the kinetic coupled gravity theories have been received specific attention
\cite{Sushkov:2009, Gao:2010, Germani, Nojiri}. In these theories, a non-minimal coupling in the kinetic term of the dynamical fields is responsible for the physics of acceleration. In recent works, new models of scalar field and
gauge field inflation in the context of kinetic coupled gravity are introduced by the authors \cite{Darabi}. In the present work, we generalize the scalar field inflation to vector field inflation in the context of kinetic coupled gravity. In this regard, we combine the model of vector inflation with the idea of kinetic coupled gravity. In doing so, we remove the potential
term of the vector field in vector inflation models \cite{vector-inflation}
and replace the kinetic term of the vector field minimally coupled to gravity by a kinetic term which is non-minimally coupled to gravity. The non-minimal coupling is set up by considering the Einstein tensor besides the metric tensor within the kinetic term, which is called kinetic coupled gravity.

\section{\textbf{Vector Inflation}}

In this section, we briefly introduce the idea of vector inflation \cite{vector-inflation}. Consider the action of a massive vector field non-minimally coupled to gravity
\footnote{$8\pi G=1$.}%
\begin{equation}
\label{action1}
S=\int dx^{4}\sqrt{-g}\left(-\frac{R}{2}-\frac{1}{4}F_{\mu \nu }F^{\mu \nu }+%
\frac{1}{2}\left( m^{2}+\frac{R}{6}\right) A_{\mu }A^{\mu }\right),
\end{equation}%
where $F_{\mu \nu }\equiv \bigtriangledown _{\mu }A_{\nu }-\bigtriangledown
_{\nu }A_{\mu }=\partial _{\mu }A_{\nu }-\partial _{\nu }A_{\mu }$. Variation of the action with respect to $A^{\mu }$ yields the equations of motion
\begin{equation}\label{EOM}
\frac{1}{\sqrt{-g}}\frac{\partial }{\partial x^{\mu }}\left( \sqrt{-g}F^{\mu
\nu }\right) +\left( m^{2}+\frac{R}{6}\right) A^{\nu }=0.
\end{equation}%
Using the spatially flat ($k=1$) Friedmann-Robertson-Walker (FRW) metric \begin{equation}\label{metric}
ds^{2}=-dt^{2}+a^{2}(t)\delta _{ij}dx^{i}dx^{j},
\end{equation}%
the equations \eqref{EOM} take the following form 
\begin{equation}
\label{EOM-1}
\frac{1}{a^{2}}\Delta A_{0}-\left( m^{2}+\frac{R}{6}\right) A_{0}-\frac{1}{a^2%
}\partial _{i}\dot{A}_{i}=0,  
\end{equation}%
\begin{equation}
\label{EOM-2}
\ddot{A}_{i}+\frac{\dot{a}}{a}\dot{A}_{i}-\frac{1}{a^{2}}\Delta A_{i}+\left(
m^{2}+\frac{R}{6}\right) A_{i}-\partial _{i}\dot{A}_{0}-\frac{\dot{a}}{a}\partial _{i}{A}_{0}+\frac{1}{%
a^{2}}\partial _{i}\left( \partial _{k}A_{k}\right) =0,  
\end{equation}%
where $\partial _{i}\equiv \partial /\partial x^{i}$, and a dot denotes 
derivative with respect to the cosmic time $t$. One may define a scalar
which is rotation invariant 
\begin{equation}
I=A^{\alpha }A_{\alpha }=A_{0}^{2}-\frac{1}{a^{2}}A_{i}A_{i},
\end{equation}%
and introduce a new variable $B_{i}= A_{i}/a$.
If we consider the
quasi-homogeneous vector field ($\partial _{i}A_{\alpha }=0$) we immediately
find from (\ref{EOM-1}) that 
\begin{equation}
A_{0}=0,
\end{equation}%
and equation (\ref{EOM-2}) becomes
\begin{equation}\label{Bi}
\ddot{B}_{i}+3H\dot{B}_{i}+m^{2}B_{i}=0,
\end{equation}%
where $H\equiv \dot{a}/a$ is the Hubble parameter. It turns out that when $H$ is larger than the mass $m$ the fields $B_{i}$ are frozen. Therefore, the potential $%
-m^{2}A_{\mu }A^{\mu }=m^{2}B_{i}B_{i}\approx const$ is expected to drive a quasi de Sitter expansion analogous to the scalar field. 
The components of energy-momentum tensor for the homogeneous vector field $B_{i}$ in the flat Friedmann universe are obtained as 
\begin{equation}
\label{E1}
T_{0}^{0}=\frac{1}{2}\left( \dot{B}_{k}^{2}+m^{2}B_{k}^{2}\right) ,
\end{equation}%
\begin{eqnarray}
\label{E2}
T_{j}^{i} &=&\left[ -\frac{5}{6}\left( \dot{B}_{k}^{2}-m^{2}B_{k}^{2}\right)
-\frac{2}{3}H\dot{B}_{k}B_{k}-\frac{1}{3}\left( \dot{H}+3H^{2}\right)
B_{k}^{2}\right] \delta _{j}^{i}  \notag \\
&&+\dot{B}_{i}\dot{B}_{j}+H\left( \dot{B}_{i}B_{j}+\dot{B}_{j}B_{i}\right)
+\left( \dot{H}+3H^{2}-m^{2}\right) B_{i}B_{j}.
\end{eqnarray}%
The spatial part of the energy-momentum tensor contains off-diagonal components with the same order of magnitude as the diagonal components and so the isotropic
FRW metric coupled to the homogeneous vector field does not satisfy
the Einstein equations. 

To remedy this problem, one way is to consider a triplet of mutually orthogonal vector fields $%
B_{i}^{\left( a\right) }$ \cite{Armendaris}, each with the same magnitude $\left\vert B\right\vert$, as 
\begin{equation}
\label{tr1}
\sum\limits_{i}B_{i}^{\left( a\right) }B_{i}^{\left( b\right) }=\left\vert
B\right\vert ^{2}\delta _{b}^{a},
\end{equation}%
from which one obtains 
\begin{equation}
\sum\limits_{a}B_{i}^{\left( a\right) }B_{j}^{\left( a\right) }=\left\vert
B\right\vert ^{2}\delta _{j}^{i}.
\end{equation}%
Using these relations one finds from
(\ref{E1}) and (\ref{E2}) that 
\begin{equation}
T_{0}^{0}=\epsilon =\frac{3}{2}\left( \dot{B}_{k}^{2}+m^{2}B_{k}^{2}%
\right) , 
\end{equation}%
\begin{equation}
T_{j}^{i}=-p\delta _{j}^{i}=-\frac{3}{2}\left( \dot{B}%
_{k}^{2}-m^{2}B_{k}^{2}\right)\delta _{j}^{i}, 
\end{equation}%
where $B_{k}$ satisfy \eqref{Bi} and $H$ is given by 
\begin{equation}
H^{2}=4\pi \left( \dot{B}_{k}^{2}+m^{2}B_{k}^{2}\right) .
\end{equation}%
These equations are exactly the same as those of massive scalar field in
scalar field inflation \cite{Inflation-Books}. Moreover, for $\left\vert B\right\vert >1$ we have the slow-roll approximation ($\dot{B}_{k}^{2}\ll m^{2}B_{k}^{2}$) for which $ p\approx -\epsilon $ and the universe undergoes inflation. The inflation is ended once the value of $\left\vert B\right\vert $ decreases towards the Planck value. 

Another way to remedy the problem of isotropy is to consider a large ($N$)
number of randomly oriented independent non-interacting vector fields with equal masses $m$ and equal
magnitudes $B$. The spatial components of these vector fields satisfy (%
\ref{Bi}) and their total contribution to the time-time component $T_{0}^{0}$ is estimated as%
\begin{equation}
T_{0}^{0}=\varepsilon \simeq \frac{N}{2}\left( \dot{B}%
_{k}^{2}+m^{2}B_{k}^{2}\right) .
\end{equation}%
In the estimation of spatial components of the energy-momentum tensor
one finds that during inflation typical values of the off-diagonal spatial components are of order $H^{2}\sqrt{N}B^{2}$. Therefore, the isotropic inflation
becomes possible only if these components are smaller than $T_{i}^{i}\sim
T_{0}^{0}\sim H^{2}$, namely for $B<1/N^{1/4}$. 
The slow roll regime is valid only if the effective friction in (\ref{Bi}),
exceeds their mass $m$, hence inflation is ended when the slow roll regime is over at $H\simeq m.$ Considering the point that during the inflationary stage 
\begin{equation}
\label{i7}
H^{2}=\frac{8\pi }{3}\epsilon \simeq \frac{4\pi }{3}Nm^{2}B^{2},
\end{equation}%
one finds that when the field $B$ drops to $1/N^{1/2}$ it starts to
oscillate and so the inflation is really finished. Therefore, provided that
$\frac{1}{\sqrt{N}}<B<\frac{1}{\sqrt[4]{N}}$, the isotropic inflation by vector fields may take place as
\begin{equation*}
{a_{f}}\simeq{a_{i}}\exp \left( 2\pi NB_{in}^2\right) ,
\end{equation*}%
where $B_{in}$ is the initial value of vector fields. 

\section{\textbf{Vector Inflation by nonminimal coupling to gravity }} 

Let us consider the action of an Abelian gauge field $A^{\mu}$ which is nonminimally
coupled to gravity as
\begin{equation}\label{The-model2}%
S=\int
d^4x\sqrt{-{g}}\left[-\frac{R}{2}-\frac{1}{4}(\kappa g^{\rho\mu}+\alpha G^{\rho\mu})(\kappa g^{\nu\lambda}+\alpha G^{\nu\lambda})F_{\rho\lambda}F_{\mu\nu}\right]\,,
\end{equation}
where $F_{\mu\nu}$ is the field strength tensor defined by
\be\label{F-general2}%
F_{\mu\nu}=\partial_\mu A_{\nu}-\partial_{\nu}
A_{\mu}, %
\ee%
and $g^{\mu\nu}$ is the metric tensor, $G^{\mu\nu}$ is the Einstein tensor and $\kappa>0, \alpha<0$ are constant parameters. 
This action is different from \eqref{action1} in two aspects: 
\\
i) there is no potential term containing $A_{\mu}A^{\mu}$, 
\\ 
ii) the kinetic term is nonminimally coupled to gravity.
\\
The introduction of Einstein tensor in the kinetic term of gauge field is motivated by the recently proposed idea of inflation from kinetic coupled gravity where the potential term is removed in favor of a nonminimal coupling in
the kinetic term \cite{Darabi}.
We work in the temporal gauge $A_{0}=0$ and in order to respect for the homogeneity
we just allow for $t$ dependent field configurations $\textbf{A}(t)$. Moreover, considering the discussion in the previous section, and in order to respect for the isotropy, we may take a triplet of mutually orthogonal vector fields $\textbf{A}^a\,(a=1,2,3)$, or consider a large number of randomly oriented vector fields $\textbf{A}^a\,(a=1..N)$, each with the same magnitude $|\phi(t)|$ \cite{vector-inflation}. 

The energy momentum tensor is obtained by variation of the action \eqref{The-model2}
with respect to the metric $g_{\mu\nu}$ as follows
\begin{eqnarray}
T_{\alpha\beta}&=&\{F{^\mu_\alpha}F_{\mu\beta}-\frac{1}{4}g_{\alpha\beta}F_{\mu\nu}F^{\mu\nu}\}
\nonumber\\
&+&\frac{\alpha\kappa}{4}\{g_{\alpha\beta}g^{\rho\mu}G^{\nu\lambda}F_{\rho\lambda}F_{\mu\nu}
-2G^{\rho\mu}F_{\rho\beta}F_{\mu\alpha}+\tilde{g}_{\alpha\beta}R
+\tilde{g}_{\lambda\nu}\tilde{g}^{\lambda\nu}R_{\alpha\beta}
\nonumber\\
&+&\square F(\phi)g_{\alpha\beta}-\nabla_\alpha \nabla_\beta F(\phi)
-4\tilde{g}_{\lambda\beta}R^\lambda_\alpha\},
\end{eqnarray}
where $F(\phi)=-6(\frac{\dot{\phi}^2}{a^2})$ and $\tilde{g}_{\lambda\nu}=g^{\rho\mu}F_{a\rho\lambda}F^a_{\mu\nu}$.
Note that, we have ignored the terms containing $\alpha^2$ by assuming $\alpha$ to be small. This assumption is based on the reasonable hypothesis that an important contribution of Einstein tensor in the kinetic term of the vector field is limited to the early universe in order to set up an inflationary era. In explicit words, we expect that $\alpha G^{\mu\nu}$ in the kinetic term of the vector field is important only in the early universe where the
scale factor is very small and $H$ is large enough to trigger the inflation. The presence of small coefficient $\alpha$ can help to achieve this goal, such that after the inflation is finished the term $\alpha G^{\mu\nu}$ becomes completely ignorable.

For simplicity, we take FRW background metric \eqref{metric}, and consider a triplet of vector fields $\textbf{A}^a$. Then, we may cast the above energy-momentum tensor in the form of a homogenous perfect fluid
$$ T^{\mu}_{\ \nu}=diag(\rho, p, p, p)\,,$$
where
\begin{eqnarray}\label{density}
\rho=\frac{3\kappa}{2}(\frac{\dot{\phi}^2}{a^2})+\frac{\alpha\kappa}{4}
\left[-3(\frac{\dot{\phi}^2}{a^2})(\frac{2\dot{H}+3H^2}{a^2})
+9H^2(\frac{\dot{\phi}^2}{a^2})
-3(\frac{\dot{\phi}^2}{a^2})R+6(\frac{\dot{\phi}^2}{a^2})R^0_0\right],
\end{eqnarray}
\begin{eqnarray}\label{P}
p=\frac{\kappa}{2}(\frac{\dot{\phi}^2}{a^2})+\frac{\alpha\kappa}{4}
\left[3(\frac{\dot{\phi}^2}{a^2})(\frac{2\dot{H}+3H^2}{a^2})-9H^2(\frac{\dot{\phi}^2}{a^2})
-3(\frac{\dot{\phi}^2}{a^2})R+6(\frac{\dot{\phi}^2}{a^2})R^i_i-\ddot{F}(\phi)\right],
\end{eqnarray}
and
\be\label{R} %
R= 6[\dot{H}+2H^2] ,\qquad R^0_0=3[\dot{H}+H^2],\qquad R^i_i=[\dot{H}+3H^2]. %
\ee%
We may divide $\rho$ and $p$ in the following way
\be\label{rho-P-total1} %
\rho= \rho_{_{A}}+\rho_\alpha\ ,\qquad p=p_{_{A}}+p_\alpha, %
\ee%
where 
\be\label{R5} %
\rho_{_{A}}=\frac{3\kappa}{2}(\frac{\dot{\phi}^2}{a^2}),\qquad p_{_{A}}=\frac{\kappa}{2}(\frac{\dot{\phi}^2}{a^2}), %
\ee%
\be\label{R6} %
\rho_{\alpha}=\frac{\alpha\kappa}{4}
\left[{-3(\frac{\dot{\phi}^2}{a^2})(\frac{2\dot{H}+3H^2}{a^2})+9H^2(\frac{\dot{\phi}^2}{a^2})-3(\frac{\dot{\phi}^2}{a^2})R+6(\frac{\dot{\phi}^2}{a^2}})R^0_0\right], %
\ee%
\be\label{R7} %
p_{\alpha}=\frac{\alpha\kappa}{4}
\left[3(\frac{\dot{\phi}^2}{a^2})(\frac{2\dot{H}+3H^2}{a^2})-9H^2(\frac{\dot{\phi}^2}{a^2})-3(\frac{\dot{\phi}^2}{a^2})R+6(\frac{\dot{\phi}^2}{a^2}
)R^i_i-\ddot{F}(\phi)\right]. %
\ee%
The Friedmann equations are obtained as
\begin{eqnarray}\label{friedmann0}
3H^2&=\rho=(\rho_{_{A}}+\rho_\alpha), \\\nonumber  2\dot{H}+3H^2&=-p=-(p_{_{A}}+p_\alpha)\,.
\end{eqnarray}
In order to obtain a successful inflationary model we assume that the effective pressure corresponding to the Einstein tensor in the kinetic term is negative as $p_\alpha=-\rho_\alpha$. This gives the equation
\be\label{condition2} %
\frac{d^2x}{dt^2}-\frac{1}{3}(\dot{H}+3H^2)x=0, %
\ee%
where
\be\label{R8} %
x=\frac{\dot{\phi}^2}{a^2}.  %
\ee%
Using this equation of state, the Friedmann equations give rise to%
\begin{eqnarray}\label{Friedmann1E}
3H^2=\rho=(\frac{3\kappa}{2}-\frac{\alpha\kappa}{4}C)x,
\end{eqnarray}
\begin{eqnarray}\label{Friedmann2E}
\dot{H}=-\kappa x,
\end{eqnarray}
where 
\begin{eqnarray}\label{A'}
C=3(\frac{2\dot{H}+3H^2}{a^2}+3H^2).
\end{eqnarray}
By using Eq.\eqref{Friedmann2E}, we obtain
\begin{eqnarray}\label{Friedmann3E}
\frac{d}{dt}( \frac{3\kappa}{2}-\frac{\alpha\kappa}{4}
C)x+6\kappa Hx=0.
\nonumber\\
\end{eqnarray}
We define the slow-roll parameters
\be\label{epsilon-eta-rho-P1}
\varepsilon\equiv -\frac{\dot H}{H^2}=\frac32\frac{\rho+p}{\rho}\,,\qquad \eta=\varepsilon-\frac{\dot\varepsilon}{2H\varepsilon}\,,
\ee
subject to $\varepsilon,\ \eta\ll 1$. Using the Friedmann equations \eqref{friedmann0}, we obtain%
\be\label{epsilon-rho0-rho11}
\varepsilon= \frac{2\rho_{_{A}}}{\rho_{_{A}}+\rho_\alpha}\,.
\ee%
The slow-roll regime is achieved provided that $\rho_\alpha\gg \rho_{_{A}}$. This is consistent, by considering Eqs.\eqref{R5}, \eqref{R6}, with large
value of $H$ and small value of the scale factor (about
the Planck length $\ell_P$) at the beginning of inflation, such that the first term containing $\alpha(\frac{2\dot{H}+3H^2}{a^2})$ in \eqref{R6} becomes very large and results in $\rho_\alpha\gg \rho_{_{A}}$\footnote{Even, if we assume $\alpha \sim \ell_P^2$, the combination $\alpha(\frac{2\dot{H}+3H^2}{a^2})\sim({2\dot{H}+3H^2})$ is large enough to validate $\rho_\alpha\gg \rho_{_{A}}$, because of the
assumed large value of $H$.}. Since $\phi(t)$ by itself is not a scalar under the general coordinate transformations, we define a scalar over FRW
background as
\be\label{psi-def} 
\psi(t)=\frac{\phi(t)}{a(t)}.
\ee%
Moreover, we define%
\be\label{delta-def1}
\delta\equiv-\frac{\dot{\psi}}{H\psi}\,,
\ee%
where $\delta$ is related to $\varepsilon$ and $\eta$ through the equations \begin{eqnarray}
\label{epsil1}
\varepsilon&=&2+\frac{\alpha\kappa}{6}\psi^2C, \\ \label{tilde-eta1}
\eta&=&\varepsilon-\frac{\varepsilon-2}{2\varepsilon}\left[+\frac{\dot{C}}{HC}-\frac{2\delta}{H}-\frac{2\dot{\delta}}{H(1-\delta)}\right]\,,
\end{eqnarray}
and
\begin{eqnarray}
\label{Adot}
\frac{\dot{C}}{HC}\approx\varepsilon^2\approx0,
\end{eqnarray}
The slow-roll regime requires $\dot\varepsilon\sim H\varepsilon^2$ and $\eta\sim\varepsilon$, so we should demand that $\delta\sim \varepsilon^2$. Explicitly, the equations \eqref{Friedmann1E}, \eqref{Friedmann2E}, \eqref{epsil1} and \eqref{tilde-eta1} admit the solutions
\begin{eqnarray}
\label{epsilon-x1}
\varepsilon&\simeq&3\kappa\psi^2,\\
\label{delta-x1}
\delta&=&\frac{\varepsilon}{2}(\varepsilon-\eta)+\frac{1}{2H}\frac{\dot{C}}{C},%
\end{eqnarray}
where $\simeq$ means equality to first order in slow-roll parameter $\varepsilon$. 
Also, by using Eqs. \eqref{psi-def}, \eqref{R8}, \eqref{Friedmann1E}, \eqref{delta-def1} and \eqref{epsilon-x1} we get 
\begin{eqnarray}
\label{A}
C=\frac{4}{3\alpha}(\frac{3}{2}-\frac{9}{\varepsilon})\approx-\dfrac{12}{\alpha\varepsilon},
\end{eqnarray}
which has a very large positive value because of small values of $\alpha<0$ and $\varepsilon>0$.
The conservation equation is
\begin{equation}
{\dot{\rho}+3H(\rho+p)=0,}
\end{equation}
or
\begin{equation}
{H\frac{dx}{dt}+\frac{3}{2}\frac{H}{\kappa}x=0},
\end{equation}
where we have used equations \eqref{condition2}, \eqref{Friedmann3E} and \eqref{A}. So, we have
\begin{eqnarray}\label{1}
3H^2=(\frac{3\kappa}{2}-\frac{\alpha\kappa}{4}C)(\frac{\dot{\phi}}{a})^2,
\end{eqnarray} 
\begin{eqnarray}\label{2}
\dot{H}=-\kappa (\frac{\dot{\phi}}{a})^2,
\end{eqnarray} 
\begin{eqnarray}\label{3}
\ddot{\phi}+(H+\frac{3}{4\varepsilon}){\dot{\phi}}=0.
\end{eqnarray}
Note that, according to \eqref{A'} and \eqref{A}, a large value for $C$ is consistent with a large value of $H$ and a small value of $\varepsilon$. Moreover, \eqref{1} gives the following equation in the slow-roll regime
\begin{eqnarray}\label{1'}
H^2=-\frac{\alpha\kappa}{12}C(\frac{\dot{\phi}}{a})^2.
\end{eqnarray} 
Using \eqref{A}, \eqref{2}, and \eqref{1'} we obtain
\begin{eqnarray}
\dot{H}\approx-\varepsilon H^2,
\end{eqnarray}
which confirms $H\approx{const}$ in the slow-roll regime
and gives rise to the inflationary behaviour  
\begin{eqnarray}\label{a} 
a=a_0e^{Ht},
\end{eqnarray}
and
\begin{eqnarray}
\label{q} q(H)=-(1+\frac{\dot{H}}{H^2})=-(1-\varepsilon)\approx-1.
\end{eqnarray}
Using \eqref{3}, the scalar field as a function of time behaves in the following form
\begin{eqnarray}
\phi(t)=c_1+c_2 \,\exp\left[-(\frac{3}{4\varepsilon}+H)t\right],
\end{eqnarray}
where $c_1$ and $c_2$ are constants of integration. By taking $c_1=0$ and
$c_2=\phi_0$ we have
\begin{eqnarray}
\phi(t)= \phi_0\,\exp\left[-(\frac{3}{4\varepsilon}+H)t\right].
\end{eqnarray}
The inflation is finished if $\rho_\alpha\leq \rho_{_{A}}$. In fact, considering
\eqref{R6}, we find that all the terms in the definition of $\rho_\alpha$ have $a^2$ or $a^4$ in the denominator and some terms are proportional to
$H^2$. Hence, considering the fact that towards the end of inflation the scale factor expands inflationary and $H^2$ becomes small through \eqref{1'}, these terms become negligible in a fraction of a second and lead to an small $\rho_\alpha$ compared to $\rho_A$.

\section{Conclusion}

In this paper, we have combined two recently proposed ideas of {\it vector
inflation} and {\it kinetic coupled gravity}. Vector inflation can give either
a completely isotropic universe using orthogonal triplet of vector fields, or slightly anisotropic universe using $N$ randomly oriented independent
vector fields. Kinetic coupled gravity, on the other hand, has been introduced mainly to account for the current acceleration of the universe. The non-minimal kinetic coupled gravity is established by introducing the Einstein tensor besides the metric tensor within the kinetic term of the dynamical field. By combining these two interesting ideas, we have obtained an inflationary model based on a vector field without a potential term but with a kinetic term which is non-minimally coupled to gravity. We have just concentrated
on the orthogonal triplet of vector fields, to recover the full isotropy. It seems the investigation on the $N$ randomly oriented vector fields here,
literally repeats almost the similar calculations for $N$ randomly oriented vector fields in \cite{vector-inflation}.

\end{document}